\documentclass[aps,prd,floatfix,nofootinbib,showpacs,superscriptaddress]{revtex4}
\usepackage{graphicx,amsmath,amssymb,bm}
\usepackage{color}
\usepackage[utf8]{inputenc}
\definecolor{purple}{rgb}{0.5,0,0.5}
\definecolor{blue}{rgb}{0.0,0,0.9}
\usepackage[colorlinks=true, pdfstartview=FitV, citecolor= purple, linkcolor = blue,urlcolor=blue]{hyperref}
\usepackage{csquotes}
\usepackage{multirow}
\usepackage{epsfig}
\usepackage{rotating}
\usepackage{longtable} 
\usepackage{url}
\usepackage{pdflscape}
\usepackage{cancel}
\pdfoutput=1
\usepackage{csquotes}
\usepackage{placeins}
\usepackage{scalerel}

\newcommand{\ie}{{\emph i.e.,\ }}

\newcommand{\qi}{q_{i}}  
\newcommand{\qj}{q_{j}}
\newcommand{\qk}{q_{k}}

\newcommand{\nni}{n_{i}}
\newcommand{\nj}{n_{j}}
\newcommand{\nk}{n_{k}}

\newcommand{\ei}{e_i}
\newcommand{\li}{l_i}
\newcommand{\di}{d_i}
\newcommand{\ui}{u_i}
\newcommand{\hi}{\phi_i}
\newcommand{\hj}{\phi_j}
\newcommand{\hk}{\phi_k}

\newcommand{\sigman}{\Sigma n}

\newcommand{\hh}{\phi_2 }
\newcommand{\hhh}{\phi_3}

\newcommand{\gzp}{g_{Z'}}

\newcommand{\yu}{Y}

\newcommand{\gp}{g_Y}

\begin{document}

\title{Minimal $Z'$ models for flavor anomalies}
\author{Richard H. Benavides}
\email{richardbenavides@itm.edu.co}
\affiliation{ Facultad de Ciencias Exactas y Aplicadas, Instituto Tecnol\'ogico Metropolitano,
Calle 73 No 76 A - 354 , V\'ia el Volador, Medell\'in, Colombia}
\author{Luis Muñoz}
\email{luismunoz@itm.edu.co}
\affiliation{ Facultad de Ciencias Exactas y Aplicadas, Instituto Tecnol\'ogico Metropolitano,
Calle 73 No 76 A - 354 , V\'ia el Volador, Medell\'in, Colombia}
\author{William A. Ponce}
\email{william.ponce@udea.edu.co}
\affiliation{Instituto de F\'isica, Universidad de Antioquia, Calle 70 No.~52-21, Apartado A\'ereo 1226, Medell\'in, Colombia}
\author{Oscar Rodr\'iguez}
\email{oscara.rodriguez@udea.edu.co}
\affiliation{ Facultad de Ciencias Exactas y Aplicadas, Instituto Tecnol\'ogico Metropolitano,
Calle 73 No 76 A - 354 , V\'ia el Volador, Medell\'in, Colombia}
\affiliation{Instituto de F\'isica, Universidad de Antioquia, Calle 70 No.~52-21, Apartado A\'ereo 1226, Medell\'in, Colombia}
\author{Eduardo Rojas}
\email{eduro4000@gmail.com}
\affiliation{Departamento de F\'\i sica, Universidad de Nari\~no, 
A.A. 1175, San Juan de Pasto, Colombia}

\begin{abstract}
By allowing gauge anomaly cancellation between fermions in different families
we find a non-universal solution for a  $Z'$ family of models with the same content of fermions of the standard model plus three right-handed neutrinos.  
We also impose constraints from the Yukawa interaction terms in such a way that at the end we obtain a solution with six free parameters. 
Our solution contains as particular cases well-known models in the literature.  
As an application, we report a model that evades LHC constraints, flavor changing neutral currents and low energy constraints. Simultaneously, 
the model is able to explain the flavor anomalies in the Wilson coefficients $C_9(\mu)$ and $C_{10}(\mu)$ without modifying the corresponding
Wilson coefficients for the first family.
In our approach, this procedure is always possible for $Z'$ masses smaller than $\sim 2.5$~TeV.
\end{abstract}

\maketitle

\section{Introduction  \label{intro}}

In recent years, 
experimental anomalies in the LHCb and in low-energy experiments~\cite{Pohl:2010zza,Aaij:2013qta,Krasznahorkay:2015iga,Heister:2016stz} have generated some theoretical 
speculation about the posibility that these results  constitute  a manifestation 
of  physics beyond the standard model~(SM).
A number of anomalies in semileptonic $B$ decays have been reported 
by the  LHCb collaboration and other experiments~\cite{Aaij:2014pli,Aaij:2014ora,Aaij:2013qta,Aaij:2015oid,Wehle:2016yoi,Aaij:2013aln,Aaij:2015esa}, 
finding various deviations from
their predicted values in the SM.
Even though the experimental results are not conclusive yet, 
the global fits improve for models where the new physics
contributions to the Wilson coefficient $C_9^{\mu}$ decrease it by a quarter of the SM prediction~\cite{Vicente:2018xbv}.  
Because the only lepton in the associated   Wilson operator is the muon field,  one of the preferred  
theoretical frameworks to explain these 
anomalies are the non-universal
models~\cite{Allanach:2015gkd,Altmannshofer:2016jzy,Ellis:2017nrp,Baek:2017sew,Bian:2017rpg,Dalchenko:2017shg,Faisel:2017glo,Alok:2017jgr,Ellis:2018xal}, 
for which the electroweak~(EW) parameters and quantum numbers are family dependent.
In general, non-universal models are restricted severely by flavor changing neutral currents~(FCNC); however, 
as it is well-known~\cite{Barger:2009qs}, we can get rid of these problems by guaranteeing that the gauge couplings of the new 
physics to the left-handed down-type quarks become identical~(We do not know anything about the mixing of the right-handed quarks so that we
can assume a diagonal matrix. That result quite useful to avoid further constraints on the $Z'$ charges).  
That is particularly important for the first and second generation.

The best-known non-universal EW extensions of the SM  
correspond to the so-called 331 models; however, simpler solutions can be built 
by restricting the additional EW sector to an abelian $U(1)$  gauge symmetry with the same fermion content of
the SM plus right-handed neutrinos. As we will show, these minimal  solutions are able to explain these anomalies 
without increasing the number of new fields and parameters.
These EW extensions   are known 
as  minimal models~\cite{He:1990pn,He:1991qd,Appelquist:2002mw,Carena:2004xs,Langacker:2008yv,Salvioni:2009jp,Crivellin:2015lwa,Ma:2016zod,Kownacki:2016pmx,Tang:2017gkz,Bandyopadhyay:2018cwu,Arcadi:2018tly},
and constitute the simplest EW extension of the SM. 
The best-known example is the left-right symmetric~(LRS) model, which has universal 
EW charges for the three families and its content of fermions excess 
the SM one by a right-handed neutrino in every family. 
Earlier in the nineties, several works pointed out the non-fundamental 
character of the universality of the EW charges~\cite{Pisano:1991ee,Frampton:1992wt,Montero:1992jk,Foot:1992rh,Foot:1994ym,Ozer:1995xi,Ponce:2001jn,Ponce:2002sg,Okada:2015bxa,Cao:2016uur,Queiroz:2016gif,Blandon:2018lca}.  
This was motivated by EW models based on string theory which, in most of the cases, result to be non-universal~\cite{Langacker:2008yv}.  
A general solution to the gauge anomalies involves a cubic Diophantine equation~\cite{Batra:2005rh}; however, 
it is possible to find solutions with continuous parameters, which turn out quite useful to build benchmark models.

A lot  of phenomenology has been based on the
minimal models~\cite{Appelquist:2002mw,Almeida:2004hj,Gauld:2013qba,Salvioni:2009jp,Crivellin:2016ejn,Altmannshofer:2016jzy,Kaneta:2016uyt,Biswas:2016yjr,Elahi:2017ppe,Asai:2017ryy,
Baek:2017sew,Chen:2017usq,Biswas:2017ait,Duan:2017qwj,Ellis:2017nrp,Bian:2017rpg,Ekstedt:2017tbo,Cao:2017sju,Lozano:2018esg,Gninenko:2018tlp},
in spite of it, most of these analysis make use of some few well-known EW charge assignments 
leaving aside other possible solutions to the gauge anomaly equations with the same content of fermions. 
A first step to know the full set of solutions was given in our previous work~\cite{Benavides:2016utf}, 
where we assume two identical families and the non-universality show up only in the  third generation.  
In the present manuscript, we allow non-universal charges for  leptons and quarks 
in the three families, which result quite convenient in the study of the LHCb anomalies. 
Under some reasonable assumptions, many of these models are able to evade the FCNC constraints.

The paper is organized as follows: in Section~\ref{sec:2} we derive 
the general expressions for the   chiral charges of the models.
  In Section \ref{sec:5} we derive the  95\% C.L. allowed 
 limits on the model parameters  by the most recent
 LHC data  and the corresponding limits by the  low energy EW data. 
  Section \ref{sec:6} summarizes our conclusions.

\section{The $SU(2)_L\otimes U(1)\otimes U(1)'$ gauge symmetry}
\label{sec:2}

The  aim  of the present work  is to build the most general parameterization
for the minimal EW extension of the  SM, limiting ourselves 
to the SM fermions plus right-handed neutrinos.
In order to accomplish our purpose it is necessary to avoid the hypothesis of universality; 
with this in mind, let us consider 
the gauge group $SU(2)\otimes U(1)\otimes U(1)'$ as a non-universal 
anomaly-free extension of the EW sector of the SM. 


In what follows $T_{1L}$, $T_{2L}$ and  $T_{3L}$ denote the generators of $SU(2)_L$, while $\yu$ and $Q_{Z'}$ denote the generators of $U(1)$ and $U(1)'$,
respectively. 
The covariant derivative for our model is given by~\cite{Ponce:1987wb}
\begin{align}\label{Eq3}
D_{\mu} = \partial_{\mu}-ig\overrightarrow{T}_{L}\cdot\overrightarrow{A}_{\mu}-i\gp Y B_{Y\mu}-i\gzp Q_{Z'} Z^{\prime}_{\mu},
\end{align} 
where  $g$, $g_{Y}$ and $\gzp$ are  the gauge  couplings associated  with the $SU(2)_L$, $U(1)$ and $U(1)'$ gauge groups, respectively,
and  $\overrightarrow{A}_{\mu}$, $B_{Y\mu}$ and $Z_{\mu}^{\prime}$ stand for the corresponding  gauge fields.

In order to find the most general solution to gauge anomaly cancellation, all families have different quantum numbers, because of this, at least two Higgs doublets are required 
in order to give masses to the three families, so:

\begin{align}
\left\langle\Phi_i\right\rangle^T=(0,v_i/\sqrt{2}),\hspace{0.5cm} i=1,2. 
\end{align}

At this stage, it is important to stress that  we do not intend to report a model, instead our purpose is to show a general solution to the anomaly cancellation equations. We added two Higgs doublets since it represents the minimal scalar field content in order to have Yukawa couplings for a non-universal $Z'$ gauge boson.  In our solution, every set of parameters represents a possible electroweak model. For every choice of the $Z'$ charges it is possible to choose additional scalars in order to reproduce the mixing angles in the lepton and quark sectors.
From general grounds, with the Higgs structure of our model it is possible to generate mass matrices with four texture zeros in the lepton and quark sectors. That is possible  since that in our solution two families couple to a single Higgs doublet and just one of the families couples to a different scalar doublet.  It is well-known that even  mass matrices with five texture zeros  are able  to generate the  mixing matrices  for the lepton and quark sectors~\cite{Ludl:2015lta}.
Thus in principle it is not forbidden for  four texture zero mass matrices to generenate the  CKM and PMNS mixings. Any case, as we mentioned above, for a particular choice of the $Z'$ charges there is possible to add new scalars if needed.

\subsection{Gauge anomaly cancellation}
\label{sec:3}
For the $SU(2)_L\otimes U(1)\otimes U(1)'$ symmetry  with the particle content shown in table~\ref{tab:pcontent}, the non-trivial gauge anomaly equations are:
\begin{table}
{\begin{tabular}{|c|c|c|c|c|c|}
\hline  
Particles & Spin &$SU(3)_C$ &$SU(2)_L$ &$U(1)_Y$ &    $U(1)'$\\
\hline
$l_{Li}$  & 1/2 & 1       &  2       &   -1/2  &    $l_i$    \\  
$e_{Ri}$  & 1/2& 1        &  1       &   -1    &    $e_i$    \\
$\nu_{Ri}$& 1/2& 1        &  1       &    0    &    $\nni$   \\
$q_{Li}$  & 1/2& 3        &  2       &   1/6   &    $q_i$     \\  
$u_{Ri}$  & 1/2& 3        &  1       &   2/3   &    $u_i$     \\
$d_{Ri}$  & 1/2& 3        &  1       &   -1/3  &    $d_i$     \\
$\Phi_i$  & 0  & 1        &  2       &   1/2   &    $\hi$      \\
\hline
\end{tabular}
\caption{Particle content. The subindex  $i=1,2,3$ stand for the family number in the interaction basis. 
In our solution $\hh = \hhh$ in such a way that only two Higgs doublets are needed.
However, sometimes we keep the notation $\hi $, which is quite convenient for notation purposes.  }
\label{tab:pcontent}
}
\end{table}

\begin{widetext}
\begin{align}\label{Eq4}
[SU(2)]^2U(1)' :\hspace{0.2cm}&0=\Sigma q+\frac{1}{3}\Sigma l,\notag\\
[SU(3)]^2U(1)':\hspace{0.2cm}&0=2\Sigma q-\Sigma u-\Sigma d,\notag\\
[\text{grav}]^2U(1)':\hspace{0.2cm}&0=6\Sigma q-3(\Sigma u+\Sigma d)+2\Sigma l-\sigman-\Sigma e,\notag\\
[U(1)]^2U(1)':\hspace{0.2cm}& 0=
 \frac{1}{3}\Sigma q
-\frac{8}{3}\Sigma u
-\frac{2}{3}\Sigma d 
+\Sigma l-2\Sigma e,
\notag\\
U(1)[U(1)']^2:\hspace{0.2cm}&
0=\Sigma q^2
-2\Sigma u^2+\Sigma d^2
-\Sigma l^2+\Sigma e^2,\notag\\
[U(1)']^3:\hspace{0.2cm} & 
0=6\Sigma q^3-3(\Sigma u^3+\Sigma d^3)+2\Sigma l^3-\sigman^3-\Sigma e^3, 
\end{align} 
\end{widetext}
where $\Sigma f=f_1+f_2+f_3$.  We  also take into account the constraints coming from the Yukawa couplings: 
\begin{align}
\mathcal{L}_Y  \supset\ 
& \overline{l}_{1_L}\tilde{\Phi}_1\nu_{1_R}+\overline{l}_{1L}\Phi_1e_{1_R}+\overline{q}_{1_L}\tilde{\Phi}_1u_{1_R}+\overline{q}_{1_L}\Phi_1d_{1_R}+\notag\\
&\overline{l}_{2_L}\tilde{\Phi}_2\nu_{2_R}+\overline{l}_{2L}\Phi_2e_{2_R}+\overline{q}_{2_L}\tilde{\Phi}_2u_{2_R}+\overline{q}_{2_L}\Phi_2d_{2_R}+\notag\\
&\overline{l}_{3_L}\tilde{\Phi}_2\nu_{3_R}+\overline{l}_{3_L}\Phi_2e_{3_R}+\overline{q}_{3_L}\tilde{\Phi}_2u_{3_R}+\overline{q}_{3_L}\Phi_2d_{3_R}+\text{h.c}.
\end{align}

The corresponding  constraints  coming from  the  terms in the above Lagrangian are~(where $\hh=\hhh$): 

\begin{align}\label{Eq5}
0=\ei  - \li+\hi  , \notag\\
0=\nni - \li-\hi  , \notag\\
0=\di  - \qi+\hi , \notag\\
0=\ui  - \qi-\hi .  
\end{align}
The solution to the gauge anomaly  equations~(\ref{Eq4}) and the constraints from the Yukawa interaction terms~(\ref{Eq5}) 
corresponds to the charges shown in 
table~\ref{tab:scenario-c}~(there are six solutions corresponding to the
permutations between the indices ijk).
In general, every one of these solutions depends on six parameters, $(q_i,\nni)$,  with $i=1,2,3$,  corresponding to the $Z'$ charges
for the quark doublet and the right-handed neutrino in every generation, respectively.  
By removing the constraint $\hj=\hk$ there are two additional solutions which will
be reported elsewhere since they do not fit well the flavor anomalies. 

\begin{table}[h!]
{\begin{tabular}{|c|r|}
\hline  
$f$       &$\epsilon^{Z'}(f)$ \hspace{1.2cm}  \\            
\hline
\hline
$l_i$     &$-3\qi$\\
$e_i$     &$-\nni-6\qi$\\
$u_i$     &$+\nni + 4\qi$\\
$d_i$     &$-\nni-2\qi$\\
\hline
 $l_j$     &$+\frac{1}{2}[\nj - \nk -3(\qj + \qk)]$ \\
$e_j$     &$-\nk -3(\qj+\qk)$\\
$u_j$     &$+\frac{1}{2}(\nj + \nk + 5\qj + 3\qk )$\\
$d_j$     &$-\frac{1}{2}(\nj + \nk + \qj + 3\qk )$\\
\hline
$l_k$     &$+\frac{1}{2}[-\nj + \nk -3(\qj + \qk)]$ \\
$e_k$     &$-\nj -3(\qj+\qk)$\\
$u_k$     &$+\frac{1}{2}(\nj + \nk + 3\qj + 5\qk )$\\
$d_k$     &$-\frac{1}{2}(\nj + \nk + 3\qj + \qk )$\\
\hline
\end{tabular}
}
\caption{The $Z'$ couplings for the  Higgs doublets $\Phi_i$ and $\Phi_j$ are
$\hi = \nni + 3\qi$ and $\hj =\hk=\frac{1}{2}[\nj + \nk +3(\qj + \qk)]$,
respectively. The higgs field  $\hi$ couples to fermions in the $i$-th family. The integers  $ijk$ are a permutation of $123$.  }
\label{tab:scenario-c}
\end{table}
By setting $(\nj-\nk)/2=L_{i}=-L_{k}= 1$, $\nk=-1$ and $\qi=\qj=\qk=\nni=0$,  from  this solution we can obtain the model $L_j -L_k$~\cite{He:1990pn} 
where $L_i$ is $1$ for the leptons in the $i$-th family and zero otherwise. From these solutions, 
the most known model is the  $L_\mu -L_\tau$  model, which has been  widely used to explain the $g-2$ anomaly~\cite{Biswas:2016yjr}.   	


\section{Mixing matrices for non-universal $Z'$ models}
Since the SM is universal there is no problem with the quantum numbers
to generate the mass matrices for the quark and lepton sectors, 
the same is true for electroweak extensions of the standard model
with universal couplings; however,  non-universal models require 
additional scalars to generate the right mixing for the SM fermions.  
\subsection{Models with a non-universal right-handed sector}
By setting in table~(\ref{tab:scenario-c}) $q_i= q_j = q_k $ and $ n_j = n_k $ the charges of the 
left-handed fermions become universal, while the right-handed charges are not.
This model could be useful  since the non-universal sector is singlet under $SU(2)_L$, 
hence, we can avoid phenomenological constraints by chosen the right-handed mixing in a convenient way. 
The Yukawa interaction terms can be chosen as:
\begin{align}
 \mathcal{L}=
 \begin{pmatrix}
 \bar{q}_{Li}^T, &   \bar{q}_{Lj}^T, &  \bar{q}_{Lk}^T 
 \end{pmatrix}
\begin{pmatrix}
y_{ii}^u\tilde\Phi_i  & y_{ij}^u\tilde\Phi_j & y_{ik}^u\tilde\Phi_j  \\
y_{ji}^u\tilde\Phi_i  & y_{jj}^u\tilde\Phi_j & y_{jk}^u\tilde\Phi_j  \\
y_{ki}^u\tilde\Phi_i  & y_{kj}^u\tilde\Phi_j & y_{kk}^u\tilde\Phi_j  \\
\end{pmatrix}
 \begin{pmatrix}
 u_{Ri} \\ 
 u_{Rj} \\
 u_{Rk} 
 \end{pmatrix}\\
 +
 \begin{pmatrix}
 \bar{q}_{Li}^T, &   \bar{q}_{Lj}^T, &  \bar{q}_{Lk}^T 
 \end{pmatrix}
\begin{pmatrix}
y_{ii}^d\Phi_i  & y_{ij}^d\Phi_j & y_{ik}^d\Phi_j  \\
y_{ji}^d\Phi_i  & y_{jj}^d\Phi_j & y_{jk}^d\Phi_j  \\
y_{ki}^d\Phi_i  & y_{kj}^d\Phi_j & y_{kk}^d\Phi_j  \\
\end{pmatrix}
 \begin{pmatrix}
 d_{Ri} \\ 
 d_{Rj} \\
 d_{Rk} 
 \end{pmatrix}.
\end{align}
The Higgs charges under the new $U(1)'$ are
$Q_{Z'}(\Phi_i)= n_i+3q_i$ and $Q_{Z'}(\Phi_j)= n_j+3q_i$.
This model avoids flavor changing neutral currents in the quark sector  associated with non-universal 
left-handed couplings (non-universal right-handed couplings are not a problem because in these cases  the mixing of the right-handed components 
is not determined by the model and can be chosen in a convenient way.). 
This model has three free parameters which are enough for several applications.  With  $\Phi_i$ and $\Phi_j$,   we can also 
give mass to the lepton sector. So, this model only needs two Higgs doublets to give mass to all standard model fermions.

\subsection{Mixing matrices for 2+1 models}
The 2+1 models have  identical  $U(1)'$ charges for the $k,j$ fermion families but allow different charges for the $i$ family,
for these models it is possible to generate the CKM mixing matrix by adding 
two additional Higgs doublets, $H^u$ and $H^d$,  coupling to the quark sector
in a procedure  similar to that outlined in reference~\cite{Bian:2017rpg}.
A similar treatment is possible in the lepton sector by adding another 
couple of Higgs doublets.
It is important to notice that one or several scalar 
fields can acquire a non-zero vacuum expectation value to
break the $U(1)'$ symmetry, so we don't expect a proliferation of Goldstone bosons.
In order to implement the 2+1 models we impose 
the conditions\footnote{Notice that the unique difference respect to the models in the previous section is the condition $q_i=q_j$} $ q_j = q_k $ and $ n_j = n_k $ 
to the  $Z'$ charges in table~(\ref{tab:scenario-c})  in such a way that the  families $ j $ and $ k $ will have identical charges.  
\begin{align}
\mathcal{L}=
 \begin{pmatrix}
 \bar{q}_{Li}^T, &   \bar{q}_{Lj}^T, &  \bar{q}_{Lk}^T 
 \end{pmatrix}
\begin{pmatrix}
 y_{ii}^{u}\tilde\Phi_{i} & h_{ij}^u  \tilde H^u      & h_{ik}^u   \tilde  H^u\\
   0                &y_{jj}^{u} \tilde\Phi_{j}  & y_{jk}^{u} \tilde \Phi_{j}\\
   0                &y_{kj}^{u} \tilde\Phi_{j}  & y_{kk}^{u} \tilde \Phi_{j}
\end{pmatrix}
 \begin{pmatrix}
 u_{Ri} \\ 
 u_{Rj} \\
 u_{Rk} 
 \end{pmatrix}
 \notag\\
 +
  \begin{pmatrix}
 \bar{q}_{Li}^T, &   \bar{q}_{Lj}^T, &  \bar{q}_{Lk}^T 
 \end{pmatrix}
\begin{pmatrix}
 y_{ii}^{d}\Phi_{i} &  0 & 0\\
  h_{ji}^d   H^d &y_{jj}^{d} \Phi_{j}  & y_{jk}^{d} \Phi_{j}\\
  h_{ki}^d   H^d &y_{kj}^{d} \Phi_{j}  & y_{kk}^{d} \Phi_{j}
\end{pmatrix}
 \begin{pmatrix}
 d_{Ri} \\ 
 d_{Rj} \\
 d_{Rk} 
 \end{pmatrix}.
 \end{align}
  
  The $Z'$ charges of the additional Higgs doublets are  $Q_{Z'}(H^u)= (n_j+4q_j-q_i)$ and $Q_{Z'}(H^d)= n_i+2q_i +q_j$. 
According to reference~\cite{Bian:2017rpg} these textures for the quark mass matrices are enough to generate the CKM mass matrix.   
 By Proceeding similarly   in the lepton sector, assuming Dirac masses for the neutrinos, 
it is possible to generate the PMNS matrix adding two  Higgs doublets $H^\nu$ and $H^e$ with $Z'$ charges
$Q_{Z'}(H^\nu)= n_j + 3 q_i$  and $Q_{Z'}(H^e)= n_i + 6q_i -3q_j$, respectively.
There is also possible to work with Majorana masses under the same assumptions~\cite{Bian:2017rpg}.
There are other ways to couple additional scalars to generate the CKM; however, we aim to exemplify the procedure.  
\section{Flavor anomalies and the electroweak constraints}
\label{sec:5}

Part of the aim of this work is to show that 
it is possible to adjust the flavor anomalies by 
minimal $Z'$ models.
In order to demonstrate this statement, we carry out a $\chi^2$ analysis including the most relevant constraints on the $Z'$  parameter space.  
 For models with axial couplings to the electron different
from zero \ie $\epsilon_{L}^{Z'}-\epsilon_{R}^{Z'}\ne 0$,  important
constraints come from  parity-violation experiments   
which result
from the measurements of the weak charges of the cesium~\cite{Patrignani:2016xqp,Wood:1997zq,Guena:2004sq},  
the electron~\cite{Patrignani:2016xqp,Anthony:2005pm} and the proton~\cite{Androic:2018kni,Patrignani:2016xqp,Androic:2013rhu}. 
Another constraint that only involves left-handed chiral charges derives from the CKM unitarity~\cite{Marciano:1987ja,Buras:2013dea}.
This constraint is important since it applies even for models with zero couplings to the quarks. 


The $C_9$ and $C_{10}$ observables, which are involved in the 
recent discussions about the LHCb anomalies~\cite{Aaij:2014pli,Aaij:2014ora,Aaij:2013qta,Aaij:2015oid,Wehle:2016yoi,Aaij:2013aln,Aaij:2015esa},
have a value different from zero in the SM; our purpose is to include in the analysis the corresponding  corrections 
to these coefficients due to the interaction of the SM fermions  with a $Z'$ gauge boson. These shifts are denoted by $C_{9}^{ \text{NP}}$ and 
$C_{10}^{ \text{NP}}$ and  are expect to be zero in the SM as indicated in table~(\ref{tab:weakq}).

\begin{widetext}
 \begin{table}
 {\begin{tabular}{|c|c|c|c|}
 \hline  
 $\mathcal{O}$                    &Value~\cite{Androic:2018kni,Patrignani:2016xqp,Altmannshofer:2017fio}  & SM prediction $\mathcal{O}_{\text{SM}}$ ~\cite{Patrignani:2016xqp}       & $\Delta \mathcal{O}=\mathcal{O}-\mathcal{O}_{\text{SM}}$\\ 
 \hline 
 $Q_W(p)$                         &$0.0719\pm 0.0045$       &$0.0708\pm 0.0003$ &  $4\left(\frac{M_{Z}}{g_{1}M_{Z'}}\right)^2 \Delta_A^{ee}\left(2\Delta_{V}^{uu}+\Delta_{V}^{dd}\right) $\\            
  \hline
 $Q_W(\text{\text{Cs}})$          &$-72.62\pm 0.43$       &$-73.25\pm 0.02$   & $Z\Delta Q_W(p)+N\Delta Q_W(n)$ \\
 \hline
 $Q_W(e)$                         &$-0.0403\pm 0.0053$    &$-0.0473\pm 0.0003$& $4\left(\frac{M_{Z}}{g_{1}M_{Z'}}\right)^2 \Delta_A^{ee}\Delta_{V}^{ee}$\\
 \hline 
$1-\sum_{q=d,s,b}|V_{uq}|^2$      &$1-0.9999(6)$          &         0         & $\frac{3}{4\pi^2}\frac{M_W^2}{M_{Z'}^2}\left(\ln\frac{M_{Z'}^2}{M_W^2}\right)  \Delta_L^{\mu\mu}\left(\Delta_L^{\mu\mu}-\Delta_L^{dd}\right)    $      \\                               
 \hline
  $C_9^{\text{NP}}(\mu)$          &$-1.29^{+0.21}_{-0.20}$&         0         &  $-\frac{1}{g_{1}^2M_{Z'}^2}\frac{\Delta_L^{sb}\Delta_V^{\mu\bar{\mu}} }{V^*_{ts}V_{tb} \sin^2\theta_W}$ \\
 \hline
  $C_{10}^{\text{NP}}(\mu)$       &$+0.79^{+0.26}_{-0.24}$&         0         &  $-\frac{1}{g_{1}^2M_{Z'}^2}\frac{\Delta_L^{sb}\Delta_A^{\mu\bar{\mu}} }{V^*_{ts}V_{tb} \sin^2\theta_W}$ \\ 
\hline
  $\frac{\sigma^{\text{SM}+Z'}}{\sigma_{SM}}$& $0.83\pm 0.18$&       1         & $\frac{1+\left(1+4s_{W}^2+\Delta_V^{\mu\mu}\Delta_L^{\nu\nu}v^2/M_{Z'}^2\right)^2}{1+(1+4s_W^2)^2}-1$\\
 \hline 
 \end{tabular}

 \label{tab:weakq}	

}
\caption{ Experimental value and the new physics prediction for the shift in 
 the weak charge of the proton $Q_{W}(p)$~\cite{Androic:2018kni}, Cesium $Q_{W}(\text{Cs})$ and the electron $Q_{W}(e)$,
 owed to the interaction with the $Z'$. 
  The fourth observable is the constraint on the violation of the first-row CKM unitarity~\cite{Buras:2013dea,Patrignani:2016xqp}. Constraints on neutrino trident production
  and the limits on the Wilson coefficients $C_9$ and $C_{10}$ are also included. For the rotation from the weak basis
  to the mass eigenstates we adopt the convention~\cite{Barger:2003hg}: $\Delta^{ff}_{L,R}=\gzp\epsilon_{L,R}^{Z'}(f)$ 
  for up-type quarks, \ie  $u$, $c$, $t$, right-handed down-type quarks, \ie  $d_R$, $s_R$, $b_R$~(to avoid FCNC) and charged leptons.  
  For left-handed down-type quarks, \ie $d$, $s$ and $b$  we use  $\Delta^{fg}_{L}=\gzp\sum_{f',f''} V_{CKM}^{\dagger ff'}\epsilon_{L}^{Z'}(f')\delta_{f',f''}V_{CKM}^{f''g}$,
  and a similar expression for neutrinos but using the PMNS matrix.
It is useful to define the vector and axial expressions  $\Delta^{ff}_{V,A}=\Delta^{ff}_{R}\pm\Delta^{ff}_{L}$~\cite{Buras:2012jb}.
    The neutron weak charge $Q_{W}(n)$ is similar to that of the proton by interchanging $u\leftrightarrow d$. }
 \end{table}
\end{widetext} 

We also include constraints coming from  neutrino trident production in the scattering of muon neutrino with nuclei.
The effective Lagrangian for the new physics involved in this process is
$\mathcal{L}_{\nu_\mu\rightarrow \nu_{\mu}\mu\bar{\mu}}=-C_W\bar{\mu}\gamma^{\alpha}\mu \bar{\nu}\gamma_{\alpha}P_L\nu$,
where $C_W=\Delta^{\mu\mu}_V\Delta^{\nu\nu}_L/(2 M_{Z'}^2)$ is the Wilson coefficient at tree level.
From this result we obtain a contribution to the neutrino-nucleon scattering  like the one shown in the last row in table~\ref{tab:weakq}~\cite{Altmannshofer:2014cfa,Bian:2017rpg}.

By choosing  $(i,j,k)=(1,2,3)$ in table~\ref{tab:scenario-c} and identifying these labels with the charges of the first, second and third family, respectively,
it is possible to obtain a solution with zero couplings to the first family, \ie $q_1=n_1=0$. 
This choice has a double purpose, first of all, to avoid the strongest constraints from colliders,
which are weakened for a $Z'$ with zero couplings to the up and down quarks,  and second,
avoid contributions of the $Z'$ boson to the $C_9(e)$ and $C_{10}(e)$ coefficients.
In order to avoid FCNC, 
we also impose that the $Z'$ couplings to the left-handed down and left-handed strange be identical.

Under these restrictions and some other on the absolute value of the charges~(see the caption in table~\ref{tab:pulls}),
we found good fits for $Z'$ masses below 2.5 TeV (see table~\ref{tab:bestfit}).

The pulls of the observables in table~\ref{tab:weakq}  are shown in table~\ref{tab:pulls}. In order to avoid a best-fit
point in the non-perturbative region in the minimization of the $\chi^2$,  we restrict the absolute value of the parameters to be less than $1$ for the 
second generation and 3 for  $l_{3}$ which corresponds to the $Z'$ left-chiral  coupling to the $\tau$~(except for $q_3$  and  $l_2$  which were set at 0.6 and 1.1875, respectively, in order 
to avoid FCNC constraints and a good fit for the $C_9(\mu)$  and $C_{10}(\mu)$, simultaneously; however, other choices are possible). 
By changing these conditions  other solutions are possible; however, our aim is to show that it is possible to build a model satisfying all the constraints.   
It is important to emphasize that because the   $Z'$ couplings to the first family are zero,
there is no contribution to the weak charge of the cesium, proton, and the electron,
hence the corresponding pulls for these observables are the same as those of SM.

\begin{table}[ht]

{\begin{tabular}{| c | c | c | c | c | c |c|c|c|}
\hline
                       & \multicolumn{7}{| c |}{Pull$^i$=$\frac{\mathcal{O}_\text{exp}^i-\mathcal{O}_\text{th}^i}{\sqrt{\sigma_{\text{exp}}^{i 2}+\sigma_{\text{th}}^{i 2}}}$}&\\ \hline
$\mathcal{O}^i$        &  $Q_W(p)$  & $ Q_W(\text{Cs})$ &$Q_W(e)$ & $\text{CKM}$ &  $C_9$ & $C_{10}$ &  $\nu\text{-Trident}$ &   $\chi^2_{\text{min}}$ \\ \hline
                        &  0.244    & 1.46              & 1.38    & -1.10       & -0.575 & 0.700    & -1.00                &   7.13 \\ \hline
\end{tabular}          
\label{tab:pulls}
}
\caption{Pulls for low energy experiments in the $\chi^2$ minimization for a $M_{Z'}=2.5$~TeV.
In this analysis we identify  $i=1,2,3$ with  the first, second and third 
generation of fermions, respectively. 
The minimization was carried out by
imposing the constraints $q_1=q_2$,
and $q_1=u_1=d_1= 0$, 
in order to avoid FCNC and LHC constraints, respectively.  This choice has a double purpose 
since it forbids any contribution of the $Z'$ to $C_9(e)$ and $C_{10}(e)$ which involve fermions of the first family. 
To evade too large lepton couplings, in order to avoid non-perturbative charges, in the second and third family we restrict 
the  $Z'$ couplings of the SM fermions of the second and third
families to have an absolute value smaller than 1 and 3, respectively~(except for $q_3$  and  $l_2$  which were set at 0.6 and 1.1875, respectively, in order 
to avoid FCNC constraints and a good fit for the $C_9(\mu)$  and $C_{10}(\mu)$, simultaneously; however, other choices are possible).  
We did not impose any constraint on the right-handed neutrino couplings $\nni$ due to the
absence of constraints on these parameters. 
For the minimization of the $\chi^2$  we restrict the absolute value of the parameters to be less than $1$ for the 
second generation, and 3 for  $l_{3}$ which corresponds to the $Z'$ left-chiral  coupling to the $\tau$.
Another sets of charges are also possible by changing these constraints.}
\end{table}

\begin{table}

{
\begin{tabular}{|c|c|c|c|c|c|}
\hline  
 $M_{Z'}=2.5$~TeV        & $i=1$ & $i=2$ & $i=3$\\
\hline
$\gzp l_i$  &0 & 1.1875  & -2.9875  \\  
$\gzp e_i$  &0 & 0.3749  & -3.8001  \\
$\gzp\nni$  &0 & 2.0001  & -2.1749  \\
$\gzp\qi$   &0 & 0       &\ 0.6000  \\  
$\gzp u_i$  &0 &\ 0.8126 &\ 1.4126  \\
$\gzp d_i$  &0 & -0.8126 & -0.2126  \\ \hline
$\gzp \hi$  &0  &\multicolumn{2}{| c |}{0.8126}          \\
\hline
\end{tabular}
} 
\caption{Best fit values for the $Z'$  chiral charges of SM fermions, right-handed neutrinos
and the Higgs doublets. $i=1,2,3$ correspond to the first, second and third 
generation of fermions.}
\label{tab:bestfit}
\end{table}

\begin{figure}
\begin{center}
\centering 
\begin{tabular}{cc}
 \includegraphics[scale=0.5]{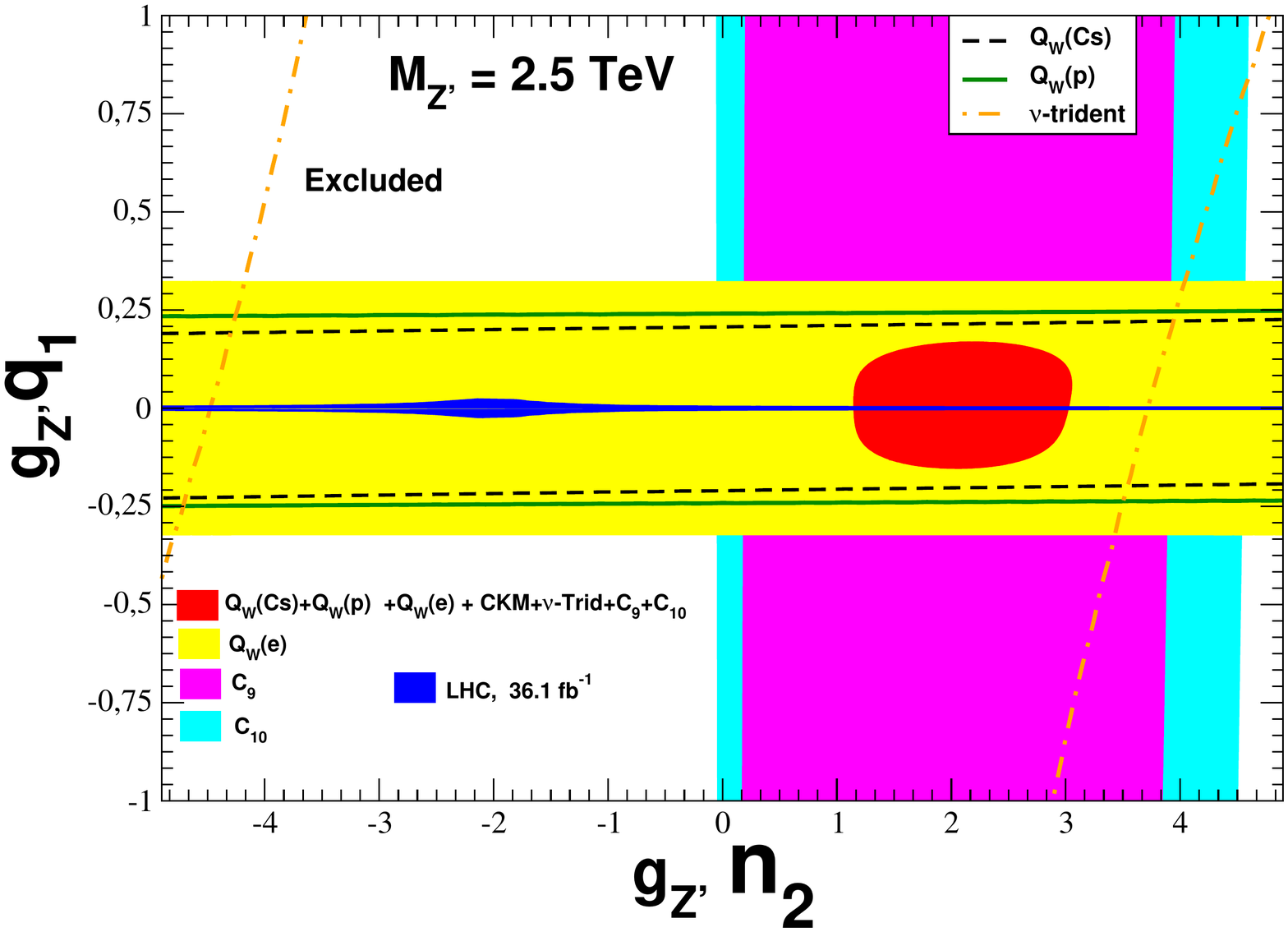}  
\end{tabular}
\end{center}
\caption{
Colored regions correspond to  the allowed parameter space  at the 95\% C.L for a $M_{Z'}=2.5$TeV.   
The region enclosed between the black-dashed lines corresponds to the 95\% C.L. allowed parameter space
by the cesium weak charge measurements~\cite{Patrignani:2016xqp,Wood:1997zq,Guena:2004sq,Erler:2009jh}. 
The yellow region corresponds to the 95\% C.L. allowed parameter space by 
 the  electron weak charge measurements in Moller scattering~\cite{Patrignani:2016xqp,Anthony:2005pm}. 
The green region corresponds to the 95\% C.L.  allowed parameter space 
by the proton weak charge measurements~\cite{Androic:2018kni}.
The region enclosed between the orange-dot-dashed lines corresponds 
to the 95\% C.L. allowed parameter space by the constraints on the violation of the first-row CKM unitarity~\cite{Marciano:1987ja,Buras:2013dea}.
By combining all the low energy data considered in our analysis  the 95\% C.L. allowed  parameter space   corresponds to the red region. 
The cyan and magenta regions correspond to the 95\% C.L.  parameter spaces consistent with the best fit values for the  $C_9$ and $C_{10}$, respectively. 
The blue region  corresponds to the 95\% C.L. parameter space allowed  by data from  
proton-proton collisions decaying to $\mu$ pairs in the ATLAS detector for an integrated luminosity of $36.1$~fb$^{-1}$
at a center of mass energy of 13TeV.
}
\label{Contours3}	
\end{figure}

In figure~\ref{Contours3} the 95\% CL allowed regions for several observables are shown. 
It is important to stress that a similar 
plot exists between any couple of parameters of the model. 
For this reason it is difficult to obtain general conclusions from this figure; however, 
the plot serves to get some idea about how each observable put constraints on the parameter space. 
These parameters, $n_2$ and $q_1$,  are important owing that they are related to the observables of our analysis.
$n_2$ appears in all the charges of the second family except in the $Z'$ coupling of the right-handed muon.
$q_1$ corresponds to the $Z'$ coupling of the left-handed up and down quark and the $Z'$ coupling of
the left-handed electron is also proportional to this parameter. 
The latter is important for the collider constraints~\cite{Erler:2011ud,Rojas:2015tqa,Rodriguez:2016cgr,Benavides:2016utf,Benavides:2018fzm}.

For the time being, the strongest constraints  come  from the   proton-proton collisions  data collected by the ATLAS experiment 
at the LHC  with an integrated luminosity  of 36.1~fb$^{-1}$  at a  center of mass energy of 13 TeV~\cite{Aaboud:2017buh}.  
In particular, 
we used the upper limits  at 95\% C.L.  on the total cross-section of the $Z'$ decaying
into dileptons~(\ie  $e^+e^-$  and  $\mu^+\mu^-$).
Figure~\ref{Contours3} shows the contours in the parameter space of the minimal models 
at 95\% C.L. for  $M_{Z'}=2.5$TeV. 
We obtain these limits from the intersection of  $\sigma^{\text{NLO}}(pp\rightarrow Z'\rightarrow l^{-}l^{+})$
with the ATLAS 95\% C.L. upper limits on the cross-section~(for additional details see  reference~\cite{Salazar:2015gxa}).
As a cross-check we calculated these limits for  the sequential SM and some $E_6$ models  finding  the same value  than that reported by 
the collaboration~\cite{Benavides:2018fzm}.
\subsection{Flavor changing neutral currents}
We assume zero mixing between the $Z$ and $Z'$ \footnote{It is true that a $\theta_{Z'-Z}$ exist owing to the existence of  Higgs sector, however, due to our particular example  in table~\ref{tab:bestfit} has zero couplings to the leptons and quarks of the first generation almost all the observables in the global analysis~\cite{Erler:2009jh}  have zero contributions  from a $Z'$ model with these couplings. Any case there is  a contribution to the mixing  from the equation 2.4 in~\cite{Erler:2009jh} but it is easier to see that this value does not surpass  $4\times 10^{-5}$.}, in such a way that  all the constraints
proportional to the $Z$-$Z'$ mixing  angle $\theta_{Z\text{-}Z'}$  in section 3.7 in the classical paper of Langacker and Plumacher~\cite{Langacker:2000ju}
 are satisfied automatically. 
The constraint coming from $\mu\text{-}e$ conversion  in a muonic atom has two contributions~(Eq.~(22) in reference~\cite{Langacker:2000ju}) , one proportional to $\theta_{Z\text{-}Z'}$  
which is proportional to $Z'$ flavor violating couplings and a contribution proportional to the $Z'$ couplings
to the up and the down  quarks which are zero in our model; therefore, the two terms are zero and satisfy the restrictions automatically.
The strongest constraints on FCNC come from the $K_0$-$\bar{K}_0$ mixing, 
and the CP violation in the Kaon system, which
are summarized in the equations 54-56  in reference~\cite{Langacker:2000ju}.
For the flavor violating $Z'$  charges of the left-handed  up and down quarks,
these constraints can be avoided by choosing the left-handed coupling $q_1$
of the quark doublet of the first generation $(u_L,d_L)$ to be identical to the corresponding 
charge $q_2$ of the doublet in the second generation $(c_L,s_L)$, 
since the $Z'$ flavor violating coupling $\Delta_L^{d,s}$ ($B^{d_L}_{1,2}$ in reference ~\cite{Langacker:2000ju}) 
are proportional to the difference between the  charges.
In order to guarantee FCNC from the $Z'$ right-handed couplings $\Delta_R^{d,s}$  
is enough with requiring a diagonal mixing between the SM right-handed fermions.  
That is possible since that in our model the parameters of the  mixing matrix of the right-handed fermions are free. 
Non-trivial constraints come from non-zero couplings$|\Delta_L^{d,b}| <6\times 10^{-8}$ 
and $|\Delta_L^{s,b}|<2\times 10^{-6}$.
In order to satisfy these constraints is enough if  $q_3<0.61$, which represents 
the $Z'$ coupling of the left-handed projection of the quarks in the third family.  
In our case we chose $q_3=0.6$. 
It is important to stress that there is a lot of 
freedom in the choice of these parameters. Our purpose
is to show that under some reasonable assumptions
it is possible to build a model.   
It is important to mention that  an update of the reference~\cite{Langacker:2000ju} is necessary in order to include the 
latest measurements of the kaon properties~\cite{Ambrosino:2006ek,Anastasi:2018qqf}.


\section{Conclusions}
\label{sec:6}

n this work we presented an anomaly-free non-universal  $Z'$ family of models, which only includes SM fermions plus right-handed neutrinos and two Higgs doublets. 
Our solutions have three families with different charges for every family, \ie the model is non-universal; however, a priori  it is not possible to identify 
one of them with a particular family in the SM;
hence, it is necessary a study of the phenomenology of all the possibilities. 

By means of an explicit example, we show that it is possible to build a model with zero couplings to the up and down quarks and in general to the fermions of the first family,  in such 
a way that the model evades collider constraints and does not contribute to the  corresponding  the Wilson coefficients  $C_9(e)$ and $C_{10}(e)$.
Simultaneously, our solution is flexible enough to accommodate  the flavor anomalies 
in the Wilson coefficients $C_9(\mu)$ and $C_{10}(\mu)$. By requiring that the left-handed couplings of the down and strange couplings be identical it is possible to avoid FCNC. 

What follows is to analyze the constraints for a $Z'$ with
strong couplings to the  $\mu$ and $\tau$ leptons but zero 
couplings to the up and down quarks~\cite{Dalchenko:2017shg}.

\section*{Acknowledgments} 
R. H. B. and L. M. thank the  ``Centro de Investigaciones  ITM''. 
We thank Financial support from ``Patrimonio Aut\'onomo Fondo Nacional de Financiamiento para la Ciencia, la Tecnolog\'ia
y la Innovaci\'on, Francisco Jos\'e de Caldas'', 
and ``Sostenibilidad-UDEA''.
This research was partly supported by the Vicerrectoría de Investigaciones, Posgrados
y Relaciones Internacionales~(VIPRI) de la Universidad de Nariño,  project numbers 1928 and 2172.

\bibliographystyle{ws-ijmpa}


%
%


\end{document}